\documentclass[]{proarticle}
\usepackage{multicol}
\usepackage{graphicx}
\usepackage{amssymb}
\usepackage{epsfig}

\def\bea{\begin{eqnarray}}
\def\eea{\end{eqnarray}}
\begin{document}
\title{Early-time cosmic dynamics in $f(R)$ and $f(|\hat\Omega|)$ extensions of Born-Infeld gravity}
\author{Andrey N. Makarenko$^{1}$, Sergei D. Odintsov$^{2,3}$,  Gonzalo J. Olmo$^{4,5}$, and Diego Rubiera-Garcia$^{6}$}
\maketitle
\address{{$^1$Tomsk State Pedagogical University, Kievskaya str., 60, 634061 Tomsk, Russia.\\}
{ $^2$Instituci\`{o} Catalana de Recerca i Estudis Avancats (ICREA),
Barcelona, Spain} \\
{$^3$Institut de Ciencies de l Espai (CSIC-IEEC), Campus UAB,}\\ {Torre
C5-Par-2a- pl, E-08193 Bellaterra (Barcelona), Spain}\\
{{ $^4$Depto. de F\'{i}sica Te\'{o}rica \& IFIC , Universidad de Valencia - CSIC}  { Burjassot 46100, Valencia, Spain }\\
{ $^5$Depto. de F\'isica, Universidade Federal da
Para\'\i ba, 58051-900 Jo\~ao Pessoa, Para\'\i ba, Brazil}}\\
{$^6$Center for Field Theory and Particle Physics, and Department of Physics, \\ Fudan University,  220 Handan Road, 200433 Shanghai, China}
}
\eads{andre@tspu.edu.ru \ , \ odintsov@ieec.uab.es \ , \ gonzalo.olmo@uv.es \ , \ drubiera@fudan.edu.cn}
\begin{abstract}
We consider two types of modifications of Born-Infeld gravity in the Palatini formulation and explore their dynamics in the early universe. One of these families considers $f(R)$ corrections to the Born-Infeld Lagrangian, which can be seen as modifications of the dynamics produced by the quantum effects of matter, while the other consists on different powers of the elementary building block of the Born-Infeld Lagrangian, which we denote by $|\hat\Omega|$.
We find that the two types of nonsingular solutions that arise in the original Born-Infeld theory are also present in these extensions, being bouncing solutions a stable and robust branch. Singular solutions with a period of approximate de Sitter inflation are found even in universes dominated by radiation.
\end{abstract}


\begin{multicols}{2}
\section{Introduction}

Understanding the dynamical laws of nature at very high energies is a challenge for theoretical physics. The idea of our Universe being born from a big bang singularity is disturbing and alternative nonsingular scenarios are desirable. In this respect, a high-energy extension of General Relativity (GR) constructed in analogy with the Born-Infeld theory of nonlinear electrodynamics \cite{BI} has been recently considered with very positive results. In this theory, formulated in a metric-affine manifold, nonsingular solutions exist that prevent the big bang. These solutions are of two types: bouncing solutions, characterized by $H=0$ and $dH/d\rho\neq 0$ at the density of the bounce, and unstable minimal volume solutions with $H=0$ and $dH/d\rho= 0$. In recent works we have studied the stability of these solutions under small perturbations of the action and also under large deformations. In this talk we summarize the main results of our analyses and the main conclusions.

\section{Perturbations of BI via $f(R)$ corrections}

\begin{figure*}[h]\center
\includegraphics[width=0.5\textwidth]{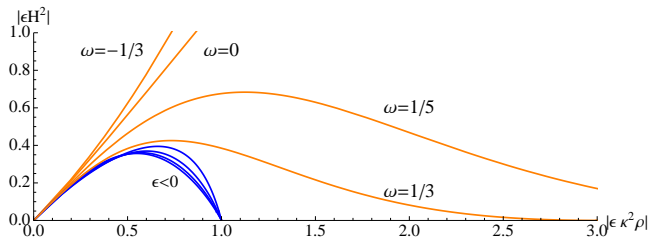}
\caption{$H^2$ in Born-Infeld theory.  }
\label{Fig:H2BI_w}
\end{figure*}

The action of the Born-Infeld theory of gravity takes the form
\begin{eqnarray}\label{eq:A0}
S_{BI}&=&\frac{1}{\kappa^2\epsilon}\int d^4x \times\\
&\times&\left[\sqrt{-|g_{\mu\nu}+\epsilon R_{\mu\nu}(\Gamma)|}-\lambda \sqrt{-|g_{\mu\nu}|}\right] \ .\nonumber
\end{eqnarray}
This action was originally introduced \cite{Deser:1998rj} and reconsidered in \cite{Banados} within the Palatini formulation, i.e., assuming that the metric and affine geometric structures are independent. For clarifications on the notation see \cite{Makarenko:2014lxa}
(for reviews on modified gravity in Palatini formulation, see \cite{Olmo:2011uz,Olmo:2012}).
It admits a power series expansion in the parameter $\epsilon$ of the form
\begin{eqnarray}
S_{BI}&\approx & \int \frac{d^4x}{2\kappa^2} \sqrt{-g} \times\\
&\times&\left[R-2\Lambda_{eff}+\frac{\epsilon R^2}{4} -\frac{\epsilon}{2} R_{\mu\nu}R^{\mu\nu}+\ldots\right] \label{eq:BI_series} \ ,\nonumber
\end{eqnarray}
where  $\Lambda_{eff}=\frac{\lambda -1}{\epsilon}$. Clearly, it recovers GR at the zeroth order and quadratic gravity with specific coefficients at the next-to-leading order. It is important to note, however, that if quantum effects of matter are taken into account, quadratic curvature corrections arise that depend on the kind and number of matter fields, which induces deviations from this effective low-energy Lagrangian. It is thus interesting to explore their potential effect on the dynamics as a test of the robustness of the predictions of this theory for the early universe. \\
In order to be as general as possible, we consider a family of extensions of the original Born-Infeld theory of the form
\begin{eqnarray}\label{eq:A1}
S_{BI-f(R)}=S_{BI}+\frac{\alpha}{2\kappa^2}\int d^4x \sqrt{-g}f(R)+S_m \ .
\end{eqnarray}
In the limit $\epsilon\to 0$ this theory can be seen as a typical $f(R)$ theory (for general review of $f(R)$ gravity, see \cite{Nojiri:2011}), whereas for $\alpha\to 0$ it recovers the original Born-Infeld theory.  The analysis and discussion of the field equations of this type of Born-Infeld-$f(R)$ theories was presented in \cite{Makarenko:2014lxa} (see also \cite{Makarenko:2014nca} and \cite{Makarenko:2014ass}). Here we simply summarize the relevant results regarding the Hubble function in cosmologies with a perfect fluid with constant equation of state $P=w\rho$. \\
In Fig.\ref{Fig:H2BI_w}, we represent the (dimensionless) Hubble function $|\epsilon| H^2$ as a function of the (dimensionless) energy density $|\epsilon| \kappa^2\rho$ in the original BI theory for different equations of state. The blue curves, which end at $|\epsilon| \kappa^2\rho=1$ represent bouncing solutions and occur for $w>-1$. The other curves are nonsingular if $w>0$ and represent unstable states of minimum volume.
\end{multicols}

\begin{figure*}[h]\center
\includegraphics[width=0.35\textwidth]{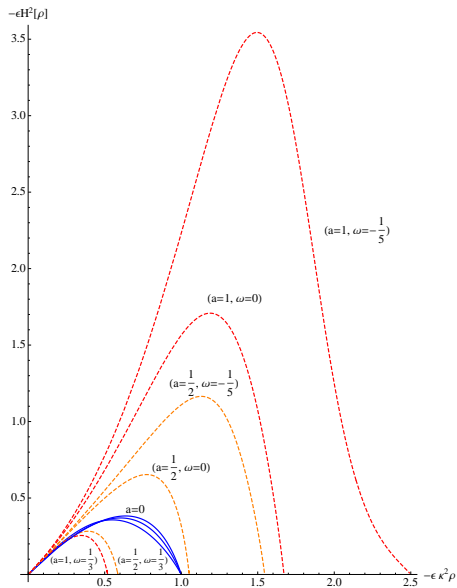}
\caption{Hubble function in BI with $f(R)=a R^2$ corrections. }
\label{Fig:Negative_Branch}
\end{figure*}

\begin{figure*}[h]\center
\includegraphics[width=0.5\textwidth]{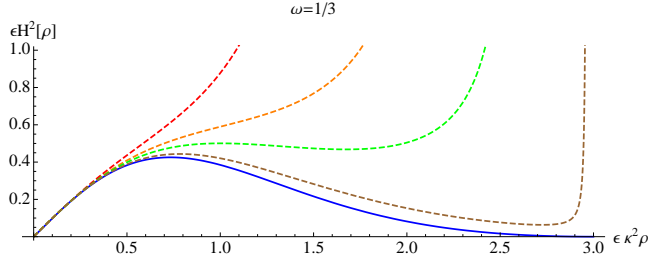}
\caption{Inflationary behavior for $a=1/3$ (dashed green curve). The solid blue curve is $a=0$. }
\label{Fig:Inflation}
\end{figure*}

In Fig.\ref{Fig:Negative_Branch}, we show that bouncing solutions exist  in the original BI theory (solid blue) and in two quadratic modifications of the form $f(R)=a R^2$, with $a=1/2$ (dashed orange) and $a=1$ (dashed red), for different equations of state ($w=-1/5,0,$ and $1/3$). The existence of a bounce appears as a robust property of the $\epsilon<0$ branch of the theory.
\\
In Fig.\ref{Fig:Inflation}, we find the Hubble function in a radiation universe   ($\omega=1/3$) in the cases $a=0$ (solid blue), $a=1/10$ (dashed brown), $a=1/3$ (dashed green), $a=1/2$ (dashed orange), and $a=1$ (dashed red). We see that a plateau follows a maximum around  $\epsilon \kappa^2\rho\approx 0.6$ in the case $a=1/3$, which could support a period of inflation generated by the radiation fluid.
\begin{multicols}{2}
The above plots put forward that the bouncing solutions  of Born-Infeld gravity are robust against modifications of the $R^2$ coefficient, whereas those in the unstable branch undergo significant changes. Remarkably, the modifications experienced by these solutions may lead to a period of de Sitter-like expansion after the big bang singularity, as is apparent from the case $a=1/3$ in Fig.\ref{Fig:Inflation}.

\section{Deformations of Born-Infeld gravity }\label{sec2}

\begin{figure*}[h]\center
\includegraphics[width=0.5\textwidth]{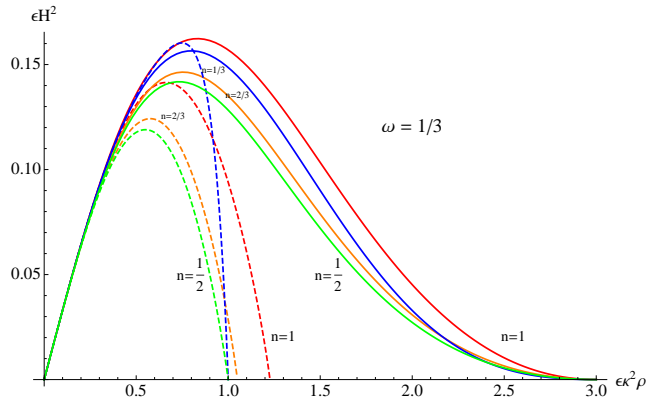}
\caption{$H^2$ for radiation in $f(|\hat\Omega|)=|\hat\Omega|^n$.}
\label{fig:H2a}
\end{figure*}

\begin{figure*}[h]\center
\includegraphics[width=0.5\textwidth]{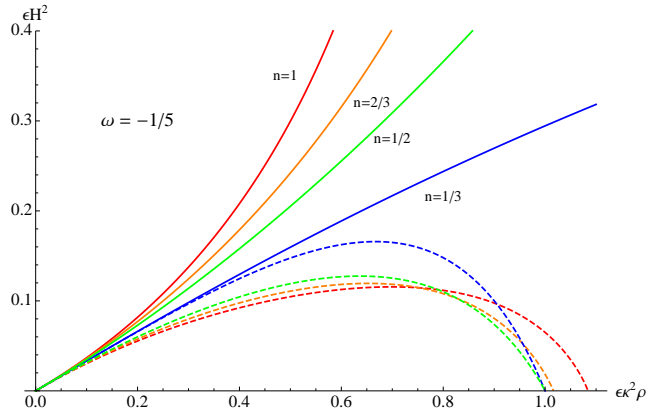}
\caption{$H^2$ for $w=-1/5$ and $n<1$.}
\label{fig:H2d}
\end{figure*}

\begin{figure*}[h]\center
\includegraphics[width=0.5\textwidth]{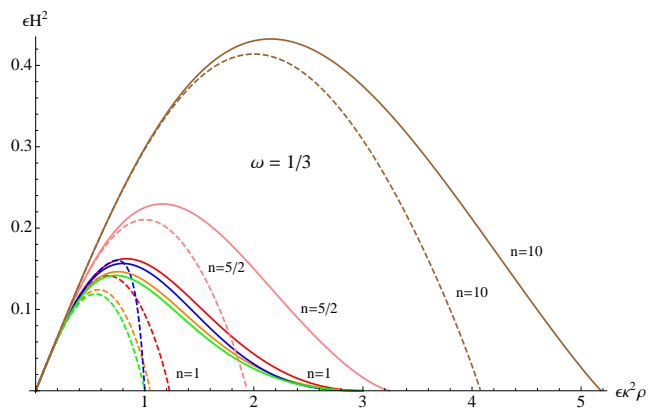}
\caption{ $H^2$ for $w=1/3$ and $n>1$.}
\label{fig:H2e}
\end{figure*}

The Lagrangian density in the action (\ref{eq:A0})  can be rewritten in a more standard form by noting that one can introduce an auxiliary metric $q_{\mu\nu}\equiv g_{\mu\nu}+\epsilon R_{\mu\nu}$ such that $q_{\mu\nu}\equiv g_{\mu\alpha}{\Omega^\alpha}_\nu$, which allows to write  (\ref{eq:A0}) as
\begin{equation}
S_{BI}=\frac{1}{\kappa^2\epsilon}\int d^4x \sqrt{-g}\left[\sqrt{|\hat\Omega|}-\lambda \right]+S_m \ .
\end{equation}
This representation suggests the following family of theories:
\begin{equation}\label{eq:f(BI)a}
S_f=\frac{1}{\kappa^2\epsilon}\int d^4x \sqrt{-g}\left[f(|\hat\Omega|)-\lambda \right]+S_m \ ,
\end{equation}
being $f(|\hat\Omega|)=|\hat\Omega|^{1/2}$ the original Born-Infeld theory. A detailed exploration of the field equations of these models and their representation in cosmological models was provided in \cite{Odintsov:2014yaa}. Here we simply discuss the results through their graphical representation.  In Fig.\ref{fig:H2a} we see that for a radiation fluid ($w=1/3$) the two types of nonsingular solutions, those of the bouncing type (dashed curves) and those representing unstable minimum volume states (solid curves), persist for small deviations of the parameter $n$ from the original Born-Infeld case $n=1/2$.
\end{multicols}
For the case  $\omega=-1/5$ (see Fig.\ref{fig:H2d}) we find that the solid lines are divergent for small values of $n$, which indicates that only the bouncing branch is able to yield non-singular solutions.
For larger values of the deformation parameter $n$ and $\omega>0$, these theories are able to yield nonsingular solutions even in the unstable branch (solid curves in Fig.\ref{fig:H2e}). In these solutions, the curve $H^2$ hits the axis
forming a non-zero angle, which indicates that they are closer to the bouncing solutions of the original Born-Infeld theory than to the unstable type, characterized by $dH/d\rho =0$ at the maximum density.
\begin{multicols}{2}

\section{Conclusions}
By exploring different extensions of the Born-Infeld theory of gravity we have been able to conclude that the avoidance of the big bang singularity by means of a bounce is a very robust result in theories of the Born-Infeld type. This property is stable against perturbations of the $R^2$ term of the low-energy expansion of the Lagrangian, which suggests that the predictions of the theory might be quite insensitive to matter loop corrections. We have also found that periods of inflationary behavior can be induced by means of $R^2$ corrections.  On the other hand, large deformations of the Lagrangian, encoded in the power of the function $|\hat\Omega|$, also preserve the bouncing branch of the original theory. The unstable solutions become more stable when $n>1$. The study of cosmological perturbations and black hole solutions in these models are subjects of future work.

\section*{Acknowledgement}
GJO is supported by the Spanish grant FIS2011-29813-C02-02, the Consolider Program CPANPHY-1205388, the JAE-doc program and  i-LINK0780 grant of the Spanish Research Council (CSIC),  and by CNPq (Brazilian agency) through project No. 301137/2014-5. A.N.M. and S.D.O. are  supported by the grant of Russian Ministry of Education and Science, project  TSPU-139 and the grant for LRSS, project No 88.2014.2. S.D.O. is supported  in part by MINECO (Spain), project FIS2010-15640.
D.R.-G. is supported by the NSFC (Chinese agency) grant No. 11305038, the Shanghai Municipal Education Commission grant for Innovative Programs No. 14ZZ001, the Thousand Young Talents Program, and Fudan University. 

\end{multicols}

\end{document}